# Enhancing Software Supply Chain Resilience: Strategy for Mitigating Software Supply Chain Security Risks and Ensuring Security Continuity in Development Lifecycle

Akinsola Ahmed, Akinde Abdullah

Department of Computer Science, Austin Peay State University, Clarksville USA.

## ABSTRACT

*This article delves into the strategic approaches and preventive measures necessary to safeguard the software supply chain against evolving threats. It aims to foster an understanding of the challenges and vulnerabilities inherent in software supply chain resilience and to promote transparency and trust in the digital infrastructure that underpins contemporary society. By examining the concept of software supply chain resilience and assessing the current state of supply chain security, the article provides a foundation for discussing strategies and practices that can mitigate security risks and ensure security continuity throughout the development lifecycle. Through this comprehensive analysis, the article contributes to the ongoing effort to strengthen the security posture of software supply chains, thereby ensuring the reliable and secure operation of digital systems in a connected world.*

## KEYWORDS

*Software Supply Chain, Security Risks, Supply Chain Resilience, Open-Source Libraries, Third-Party Components, SDLC, Security Threats, Data Protection, Malware Prevention.*

## 1. INTRODUCTION

In an era dominated by digital transformation, the software supply chain is essential to creating and implementing programs that run a networked world (Nissen and Sengupta, 2016). While code, configurations, libraries, plugins, open-source and proprietary binaries, and container dependencies make up the software supply chain (Tucci et al., 2005), Andreoli et al. (2023) observe that this connectivity leaves the software supply chain open to a wide range of security threats, from deliberate assaults to unintentional weaknesses. Consequently, vulnerable software supply chain attacks can lead to backdoor access, malware installation, application downtime, and data leakage such as passwords or private information (Ohm et al., 2020). Hence, increasing software supply chain resilience is critical as companies depend increasingly on open-source libraries, third-party components, and collaborative development methods (Linton, Boyson, and Aje, 2014).

Consequently, this article explores the strategic ideas and preventative actions required to protect the software supply chain from changing threats. Therefore, this article encompasses understanding the challenges and vulnerabilities to software supply chain resilience and assists in establishing openness and confidence in the digital infrastructure that powers modern society. Firstly, the software supply chain resilience concept and the current state of software supply chain security will be explored. Against this backdrop, strategies and practices for mitigating security risk and ensuring continuity in the development lifecycle.

 

## 2. LITERATURE REVIEW

### 2.1. Understanding Software Supply Chain Resilience

Identifying the software supply chain (SSC) is vital to comprehend software supply chain resilience. Whatever interacts with an application or contributes to its development throughout the software development life cycle (SDLC) is considered a part of the software supply chain (Ergasheva and Kruglov, 2020). As part of the SDLC, the SSC could be likened to Shylesh's (2017) outline which includes every phase and element of developing a software program or system, from inception to completion, launch, and continuous upgrades. Hence, coding, version control, testing, integration, packaging, and distribution are some of the responsibilities involved in this supply chain. From a consumer-oriented perspective, a software supply chain consists of all the components required to produce a software product, including the software developers, open-source libraries, bespoke code, and DevOps technologies (Sonatype Inc., 2018). Hence, it could be resolved that the software supply chain is a networked system of online third-party source sharing and often reflects a collaboration of several stakeholders who play a part in software development.

As a division of supply chain resilience, SSCRM (Software Supply Chain Risk Management) is the process of detecting, analyzing, and managing risks related to third-party software components and services incorporated into software offerings (Keskin et al., 2021). SSCRM entails comprehending the possible vulnerabilities that may emerge from those elements and taking actions to mitigate the risk of abuse or exposure to the software framework or consumers (Librantz et al., 2020). On the other hand, Taherdoost (2022) demonstrates that cybersecurity standards are information security standards and a set of written procedures that specify how different security measures should be implemented, managed, and kept an eye on. Cybersecurity standards are defined sets of best practices, procedures, and guidelines that ensure the safety of information systems (Srinivas, Das, and Kumar, 2019). Therefore, to reduce vulnerabilities and improve the overall resilience of the software supply chain, SSCRM and cybersecurity standards set a baseline for security measures (Cains et al., 2021).

Within the scope of Supply Chain Resilience is the development phase, a diverse strategy is used to strengthen the security of the program being developed. One of the strategies of this program is secure coding, which involves developing code that follows code security best practices; to shield and protect published code against known, unidentified, and unforeseen vulnerabilities (Zhu et al., 2014). Code reviews and static analysis are another two key strategies for enhancing software code quality and security, which can assist in identifying flaws, bugs, vulnerabilities, and code smells before they develop issues for the production environment. DevSecOpsemphasizes collaboration among development, security, and operations teams, encouraging the integration of security practices, tools, and automation from the start of development to deployment. This integration guarantees that security is not a last-minute concern but an integral and ongoing component of the software supply chain, considerably increasing its resistance to possible attacks (Andersson, Hedström and Karlsson, 2022).

### 2.2. Security Risks in Software Supply Chains

Security Magazine (2024) report indicates that software supply chain security is ranked as a key priority by 52% of chief information security officers and 70% of developers. However, the growing intricacy of contemporary software supply chains poses substantial hurdles in adequately monitoring and safeguarding each component and dependence (Chen and Wen, 2023). According to research by Chen and Wen (2023), unlike in the past when software was predominantly developed in-house by small teams, modern development practices now involve integrating open-source libraries with proprietary code across extensive development teams.

Sobb, Turnbull, and Moustafa (2020) note that this complexity is further compounded by the inclusion of numerous third-party components, such as Software as a Service (SaaS) tools, Application Programming Interfaces (APIs), and libraries, often sourced from a diverse array of vendors and providers. The fast-paced nature of software development, characterized by frequent updates and new releases, exacerbates the difficulty of tracking and securing every element of the software supply chain. As a result, security operations teams face considerable challenges in managing and ensuring the security of the software supply chain, as outlined by Charles, Emrouznejad, and Gherman (2023).

As a backdrop to insufficient visibility caused by the complexity of the software supply chain, organizations frequently struggle to maintain a full and real-time perspective of all components and processes involved (Ogheneovo, 2023). This lack of visibility often creates blind spots in which security vulnerabilities or unusual activity go undetected. Companies that do not have insight into their supply chain risk losing track of their vendor's network. Software complexity is frequently seen to be the enemy of software security. Nevertheless, the availability of diverse security standards throughout the software sector presents a substantial problem in developing consistent and generally secure methods across the supply chain (Sánchez Lasheras, Comminiello, and Krzemień, 2019) also complicated firewall settings and configurations can frequently result in an administrative headache. According to Tariq et al. (2023), the low computing capacity of many IoT devices adds to this complexity and makes advanced security solutions challenging to deploy. With a wealth of standard procedures and Supply Chain Risk Management (SCRM) systems, security teams frequently want increased visibility and safety oversight of these third-party elements.

## 2.3. Incidence of Software Supply Chain Attacks and their Impacts

One notable incident of a software supply chain attack is the MOVEit data breach, an incident that accentuates the cascading effects a breach in one component of the digital supply chain can have a disruptive effect on a multitude of stakeholders.

MOVEit, a managed file transfer software developed by Progress Software, is designed to securely exchange critical information between businesses, government agencies, and their partners. Its widespread adoption across various sectors, including healthcare, finance, and government, makes it a crucial node in the digital supply chain of many organizations. The software's role as a secure conduit for sensitive data transactions places it at the heart of organizational operations, where integrity and confidentiality are paramount.

In May 2022, it was disclosed that MOVEit had been compromised this compromise was not merely an isolated incident but a part of a larger, coordinated supply chain attack aimed at infiltrating the networks of organizations reliant on MOVEit for secure data transfer. The attackers exploited vulnerabilities within the MOVEit software to gain unauthorized access, thereby exfiltrating sensitive data being processed through the system. The breach of MOVEit software had a pronounced ripple effect, impacting a wide array of organizations that depended on the software for secure data exchange. This incident highlights the interconnected nature of modern digital ecosystems, where a single component's vulnerability can compromise the entire network's security.

In the same manner of attack, In early 2021, the cybersecurity world was rocked by the disclosure of significant vulnerabilities within the Microsoft Exchange Server software. These vulnerabilities provided threat actors with a gateway to infiltrate email servers globally, affecting a wide array of organizations worldwide. Leveraging the vulnerabilities, the threat actors established a foothold by creating a web shell, a tool that enabled them to remotely control the compromised server. This control allowed for the extraction of data, including emails and account information, and facilitated the deployment of additional malware to maintain persistence. The incident not only underscored the indispensable need for timely software patching and supply chain security but also cast a spotlight on the intricate challenges of protecting enterprise systems in an era where cyber threats are increasingly sophisticated and pervasive.

Also, in May of 2021, the ransomware attack on Colonial Pipeline, a major U.S. fuel pipeline operator, marked a significant escalation in the cyber threat landscape, particularly against critical infrastructure. Colonial Pipeline, which supplies roughly 45% of the East Coast's fuel, including gasoline, diesel, and jet fuel, was forced to halt operations due to the cyberattack. The cybercriminal group known as DarkSide got a foothold into the colonial pipeline through an exposed password for a VPN account obtained from a prior data breach on a different platform.

 SolarWinds, a key player in the service management industry, was targeted in an attack that originated in September 2019. Like every other supply chain attack, rather than attempting to infiltrate an organization's networks directly, a supply chain attack targets a third party with access to its systems. According to Oladimeji and Kerner (2023), the third-party software, in this instance the SolarWinds Orion Platform, opens a loophole that allows hackers to gain entry and spoof target organizations' users and accounts. The threat actor injected malware into the Orion system updates that were capable of reading system data and blending seamlessly with regular SolarWinds operations, making it undetectable by standard antivirus software. The fallout from this attack was significant, with approximately 18,000 SolarWinds customers inadvertently installing updates that contained malicious software. This breach allowed hackers to access customer information and conduct espionage activities against various organizations, as reported by Fortinet in 2023.

## 3. STRATEGIES FOR MITIGATING SUPPLY CHAIN SECURITY RISKS

### 3.1. Secure Development Practices

The term "secure development" describes the collection of methods, procedures, and instruments developed to find and fix security vulnerabilities in software systems early on, when it is most economical to do so (Venson, Boehm, and Clark, 2023). Consequently, static code analysis and regular, comprehensive code reviews are essential to safe development processes (Ivimey-Cook et al., 2023). As Alenezi et al. (2022) observe, these procedures provide a proactive defense against coding mistakes and security flaws by thoroughly inspecting the software code at different development lifecycle phases. For instance, in code reviews, seasoned professionals carefully evaluate the codebase for overall code quality, possible vulnerabilities, and adherence to secure coding standards (Yauney, Bartholomew, and Rich, 2021). Simultaneously, Stefanović et al. (2021) observe that static code analysis tools automatically examine the source code without running it, looking for trends that might point to coding errors or security flaws. Through the early use of these techniques, development teams may identify and address security risks before they become serious ones. This is because it aligns with the notion of proactive risk mitigation in the dynamic software development landscape, while also assisting in the building of strong and resilient software by reducing the chance of security exploits in the deployed program (Alqaradaghi, Nazir and Kozsik, 2023). However, some critics have argued that this strategy is resource, time and experience intensive, making it difficult for start-ups or small organizations to implement it successfully. Besides, there is a risk of over-dependence on standardized security protocols that may not effectively address emerging or unique threats. Besides, the security measures could impede innovation and user experience, leading to resistance from users of teams (Alqaradaghi, Nazir and Kozsik, 2023).

This strategy is supported by the 'secure by design' principle, which advocates for integrating security measures throughout the software development framework. Proponents of this theory contend that it reduces vulnerabilities as it prioritizes security from the outset, thus improving overall system resilience and minimizing the risk of exploitation. However, critics claim that this theory is not foolproof. They believe that vulnerabilities still exist due to human error and continuous threat landscapes despite best efforts (Venson, Boehm and Clark, 2023). They also argue that this theory advocates for overly stringent security measures against risks, which can hinder agility and innovation, leading to cumbersome development processes. They also note the difficulties experienced in balancing usability and security, as complex security protocols can lead to workarounds or frustrate users (Soto-Valero et al., 2021).

## 3.2. Third-Party Dependency Management

Third-party dependencies, which include libraries, frameworks, APIs, and cloud platforms, are outside parts or services that a software development project need. Evaluating third-party providers' security protocols, performance history, and adherence to industry standards are all part of a comprehensive risk assessment process and reduce the risk that might arise from dependency on other parties and empower organizations to make well-informed decisions about incorporating external components (Soto-Valero et al., 2021). The proliferation of open-source components, though cost-effective and innovative, can introduce peculiar security challenges including the potential for malicious actors to exploit common libraries, and the discovery of vulnerabilities (Shah et al., 2019). Likewise, a proactive approach to third-party dependency management allows companies to leverage on the resources and experiences of trusted third-party vendors for components like APIs, frameworks and libraries, accelerating time-to-market and reducing the burden on in-house development teams. Additionally, by depending on established third-party solutions, businesses can tap into the robust security measures implemented by these vendors like security audits, patches and regular updates, thus, reducing the risk of software vulnerability. However, critics of this strategy raise valid concerns about its drawbacks. First, they argue that over-dependence on third-party components makes companies trust and rely more on external entities, posing risks to such entities if they experience breaches or fail to maintain adequate security measures. The volume of dependencies on modern software can also make it challenging to manage and track them effectively, leading to outdated components or oversight of vulnerabilities (Castellanos Ardila, Gallina and Ul Muram, 2022). Besides, third-party components may not align properly with a company's compliance standards or security parameters, necessitating extra validation or customization efforts.

## 3.3. Software Compliance Approach

Ensuring that an organization's software licenses are utilized following the terms and conditions provided by the vendor is known as software compliance (Castellanos Ardila, Gallina and Ul Muram, 2022). Therefore, the number of licenses a user may have purchased, and the number of licenses installed on a machine must match, should a vendor do an audit. Legal and security considerations dictate that license requirements for third-party components must be followed. This entails monitoring installations and usage, maintaining accurate documentation, and comprehending the conditions governing software licenses. Software providers should monitor

International Journal on Soft Computing (IJSC) Vol.15, No.1/2, May 2024their clients to maximize software license compliance. They may accomplish this by looking at data and conducting client audits. Understanding the licensing agreements helps evaluate potential risks related to using external dependencies. Non-compliance may result in legal complications (Venson, Boehm, and Clark, 2023). Mubarkoot et al. (2022) also demonstrated that strong data management solutions are required to ensure the security and integrity of sensitive information while complying with various requirements. These solutions are stated in ISO/IEC 27001. One of the key advantages of a compliance program is that it ensures one stays within legal limitations and reduces any risks or fines.

However, this strategy has its limitations. Firstly, it focuses mostly on meeting regulatory standards than prioritizing security measures. This can create a false sense of security, as compliance does not guarantee protection against every potential threat. Secondly, compliance requirements may lag behind emerging security risks, exposing companies to new threats not covered by existing regulations (Chatterjee, Gupta and De, 2023). Besides, this approach can lead to a checkbox mentality, where businesses do not pursue stringent security practices but only aim to meet compliance requirements. Also, compliance practices vary across regions and industries, leading to inconsistencies and complexities in implementation. Lastly, compliance audits are periodical, leaving assessment gaps and vulnerabilities that may go unnoticed for some time (Valdés-Rodríguez et al., 2023).

### 3.4. Identity Access Management (IAM)

To enhance security within Identity and Access Management (IAM) systems, it is imperative to mandate the implementation of Multi-Factor Authentication (MFA). Traditionally, users accessed systems using a combination of a password or PIN and a unique identifier. However, this method often falls short in protecting against sophisticated cyber threats which is evident in the colonial pipeline attack, though the password obtained was relatively complex the system lacked another layer of authentication that could have potentially mitigated the attack (TechTarget). MFA strengthens security by requiring two or more verification factors to authenticate a user's identity. This additional layer of security significantly reduces the risk of unauthorized access, even in scenarios where user credentials might be compromised. The inclusion of a dynamically generated, one-time token, which the user accesses through a secure method, is a common and effective approach in MFA systems. By making MFA mandatory, organizations can better safeguard their critical development and deployment environments, ensuring that access is securely controlled and reducing the overall vulnerability of systems to cyber-attacks.

Advanced access control techniques like as role-based access control (RBAC), which limits network access based on an individual's function within an organization, have become commonplace (Zhang, 2022). The roles in RBAC denote the levels of network access that workers have. Restricting access permissions according to work duties through using RBAC helps reduce the possibility of unauthorized access. This approach improves security by guaranteeing that people only have the permissions required for their responsibilities in the development and deployment processes. RBAC gives both fine-grained and wide control over end users' capabilities. Users can be classified as administrators, specialists, or end users, and roles and access rights can be matched to the appropriate user (Zhang, 2022). However, this strategy has several limitations. First, IAM systems can be complex to manage and implement, requiring significant resources and expertise. This complexity can lead to mismanagement and configuration errors, potentially leading to access control issues or security vulnerabilities (Yeboah-Ofori and Islam, 2019). Secondly, the fast-paced and dynamic nature of contemporary IT environments may be challenging for IAM systems, especially for businesses with complex access requirements or frequent personnel changes. This can cause inaccuracies or delays in revoking or granting access, increasing the risk of data breaches or unauthorized access. Besides,

IAM systems may find integrating with diverse and legacy systems difficult, leading to access control inconsistencies and interoperability issues across the organization (Parker, 2023).

## 3.5. Zero-Trust Implementation

Zero-trust implementation is another approach that should be implemented across the board in the development and deployment environment to mitigate against attacks leveraging supply chain, vulnerabilities. This framework operates on the principle of "never trust, always verify," ensuring that each component within the system is authenticated and authorized before being granted access. By applying zero-trust principles, organizations can significantly reduce the attack surface by limiting access to resources to only those users and devices that meet the security criteria at any given time. This approach includes strict identity verification, micro-segmentation of networks to isolate and contain potential breaches, and continuous monitoring of network traffic and user behavior to detect and respond to anomalies in real time. Implementing zero-trust can help protect against various forms of cyber threats, including those originating from compromised third-party services and insider threats. Additionally, it supports compliance with stringent data protection regulations by providing robust mechanisms for data access control and auditability.

## 4. EVALUATING THE STRATEGIES FOR ENSURING SECURITY CONTINUITY IN THE DEVELOPMENT LIFECYCLE

### 4.1. Incidence Response Planning

Organizations are guided through eliminating the incident's source by a strong incident response strategy. Establishing reliable methods for identifying and evaluating security issues is the first step in incident response planning (Van der Kleij, Kleinhuis and Young, 2017). To quickly discover abnormalities or security breaches, entails putting advanced monitoring tools and threat detection systems into place. For instance, one of the most important steps in protecting the digital infrastructure is to use advanced threat detection technologies. Alsmadi (2023) demonstrates that these systems use artificial intelligence, machine learning, and behaviour analysis to spot anomalous patterns and possible dangers instantly. To help formulate an efficient response plan, the analysis step goes into understanding the nature and extent of the incident (Haulder, Kumar and Shiwakoti, 2019). The capacity to recover quickly and effectively should be included in this since it is a sign of a well-thought-out incident response strategy. However, it is important to note that these technologies must be monitored closely as they are exposed to risks of data breaches, theft or other malware that may affect their functionality (Alsmadi, 2023).

Consequently, the incident response plan describes how to quickly contain the problem when it is discovered to stop more harm. Impacted systems are isolated during this phase, and immediate hazards are mitigated. Eradication, the next step following containment, involves using a more permanent solution (Nyre-Yu, Gutzwiller and Caldwell, 2019). Therefore, it is necessary because removing the access points that hostile actors utilize to assault the network should be a top priority for incident response teams and be implemented to get rid of the incident's underlying cause. This includes locating and eliminating any malware, malicious code, or security holes that initially enabled the event to happen. Eradication is essential to make sure that the same catastrophe doesn't happen again. Resolving the vulnerabilities exposed during the event can entail deploying security updates, patching vulnerabilities, and deploying permanent solutions (Alsmadi, 2023).

## 4.2. Vulnerability Assessment and Management

Modern information security methods are based on vulnerability assessment, a fundamental process that identifies, categorizes and ranks vulnerabilities in computer systems, applications, and network infrastructures (Aboelfotoh&Hikal, 2019). Under this model, reactive security measures are replaced with proactive ones. Organizations can develop focused plans for risk reduction and resilience development by performing thorough assessments, which provide valuable insights into their sensitivity to possible threats.

Conversely, vulnerability management includes continuous reporting, remediation, and evaluation procedures (Walkowski et al., 2021). The first step of the assessment process involves determining and categorizing the IT infrastructure's vulnerabilities. The next step is prioritization, in which organizations may effectively allocate resources by evaluating risks according to their severity and contextual awareness. The next phase implements actions to mitigate the identified risks. Mitigation tactics try to lessen the chance of exploitation, whereas remediation strategies try to remove vulnerabilities if it is practical to do so completely.

## 4.3. Secure Coding Practices

Secure coding standards outline the procedures, choices, and practices for creating software to minimize security flaws (Meng et al., 2018). Even if prioritizing security over expediency means a delayed development process, these standards aim to direct developers in that direction. For example, "default deny" access permissions—in which users are not allowed to access sensitive resources unless they explicitly grant permission—are supported by safe coding techniques (Melara et al., 2019). The Open Web Application Security Project (OWASP) provides the top eight secure coding best practices checklist, which covers 14 categories to consider during the software development life cycle (Ramirez et al., 2020). These practices include applying security by design, maintaining solid passwords, strict access control procedures, efficient error handling and logging systems, careful system configuration, thorough threat modeling, reliable cryptography procedures, and careful input validation and output encoding. Security by design strongly emphasizes risk reduction, minimizing future technical debt, and prioritizing security throughout development (Tsoukalas et al., 2020). Password management entails imposing strict password requirements and implementing multi-factor authentication to reduce password vulnerabilities. Restricting access to sensitive data to authorized users just means implementing a "default deny" strategy. By identifying and recording software faults for later examination, error handling and logging systems reduce the impact of defects in software (Thota et al., 2020). To reduce vulnerabilities caused by out-of-date software, system configuration includes optimizing system components and ensuring frequent updates are applied. Threat modeling makes identifying and reducing attack vectors easier, whereas cryptographic techniques entail data encryption and adherence to safe essential management procedures. Input validation and output encoding also guarantee that inputs of untrusted data are vetted and encoded, lowering the risk of injection attacks.

## 4.4. Threat Modeling

Threat Modeling is identifying and ranking possible cyber threats and then implementing security measures to mitigate them (Xiong &Lagerström, 2019). To handle security concerns across domains, including software applications, networks, shared systems, IoT devices, and business processes, this approach requires cooperation between security architects, operations, and threat intelligence teams. The five main steps in the threat modeling process are application architecture diagram creation, data flow analysis threat identification, threat mitigation, and threat validation (ensure that risks are reduced) (Jbair et al., 2022). Threat modeling approaches, like STRIDE,

DREAD, PASTA, VAST, OCTAVE, and NIST threat modeling frameworks are customized to cater to distinct organizational requirements and threat situations (Vidalis, 2022). Using these approaches, cybersecurity experts may create focused security plans and strategies by evaluating risks, vulnerabilities, and their effects on business operations. It is essential to comprehend the operation of threat modeling, which entails identifying possible threat actors and estimating the harm they might do to computer programs or devices. Threat modeling assists organizations in identifying vulnerabilities and security implications at every stage of the development lifecycle by studying software architecture and business conditions (UcedaVelez& Morana, 2015). It also supports the implementation of efficient security measures and the identification of critical system components.

### 4.5 Security Testing and Quality Assurance

Software testing is essential for more than just finding bugs or improving software (Böhme, 2019). Its main goal is to reduce risk by proactively finding and fixing problems that could affect end users. An essential component of this process is quality assurance (QA), which guarantees that organizations provide their clients with the highest services (Summers, 2019). QA testing improves software product quality while strictly adhering to predetermined standards (Thörn et al., 2022). The four steps of the Plan, Do, Check, and Act cycle—also referred to as the Deming cycle or PDCA cycle—are followed in quality assurance (Realyvásquez-Vargas et al., 2018). Organizations go through these procedures repeatedly in order to assess and improve processes. In the "Do" phase, processes are developed, tested, and adjusted as needed. In the "Check" phase, processes are supervised, modified, and verified. The "Act" phase involves taking steps to promote process changes. By employing QA, one can minimize the possibility of issues in the finished product by ensuring that services are developed and executed according to the proper protocols. Various technologies are used for QA testing, including test management functional and API testing tools (Olsina et al., 2022). Certifications that conform to standards include ISO9000, Test Maturity Model (TMM), and Capability Maturity Model Integrated (CMMI). Phases of security testing ensure that security measures are in place, including evaluation, design, development, testing, deployment, and monitoring (Casola et al., 2020).

### 4.5. Security Training and Awareness Programs

Employee education on the value of alertness and proactive defense tactics is one of the main ways that security awareness training, led by Information Technology (IT) departments and cybersecurity experts, mitigates software security risks (Griffin, 2021). This training usually includes assessments and simulated scenarios to provide the knowledge and abilities to recognize and effectively address possible cyberattacks. Its main goal is to create an organizational culture of cybersecurity so that staff members are empowered to act as the first line of defense against cyberattacks. Changing employees' perspectives from reactive to proactive—instilling a sense of accountability and understanding of cybersecurity issues—is one of the main advantages of security awareness training. Organizations may considerably mitigate the likelihood of successful cyberattacks and data breaches by providing personnel with training on the most recent cyber threats and attack methods. Proficient security awareness training aids staff members in comprehending the significance of conforming to cybersecurity protocols and guidelines, consequently fortifying the organization's security stance. Security awareness training creates and preserves consumer trust and safeguards the organization's digital assets. However, businesses find implementing security awareness training programs difficult because they need more tools and knowledge to create and present engaging training materials. In the current digital era, consumers are becoming more worried about protecting their data, and businesses that show a commitment to cybersecurity stand a better chance of winning their trust and business (Wynn & Jones, 2023). Employers can show their commitment to protecting client data and upholding data availability, confidentiality, and integrity by funding security awareness training for staff (Halim et al., 2023).

## 4.6  Compliance Monitoring and Enforcement

Given the current state of data protection legislation and industry standards, vigilant compliance is required (Andrew & Baker, 2021). The stakes are higher since lawmakers and regulators threaten to charge organizations heavily for not aligning their cybersecurity and compliance efforts (Williams, 2018). Regulatory compliance guarantees adherence to the law and improves an organization's security posture by establishing a baseline of consistent minimal security requirements. Cybersecurity compliance promotes a standard approach to risk management by keeping up with the most recent laws and regulations. Its main goal is to fulfill data management and protection regulations, giving organizations a road map to follow cybersecurity best practices and reduce the risk of data breaches. In addition to guiding firms in reducing risks, compliance gives firms practical solutions for efficiently addressing expensive breaches. Compliance-related operations also maximize system dependability, resilience, and ongoing device and network monitoring (Li et al., 2023). In many industries, compliance includes financial and personal data. Consequences can follow noncompliance, especially in sectors where much sensitive data is at risk, like healthcare and banking (Coventry & Branley, 2018; Dhakad, 2023). States and nations have different requirements for compliance, and different industries have different standards. Noncompliance has consequences, from financial fines in civil court to jail time. Compliance is also essential to fostering stakeholder confidence by committing to safeguarding sensitive data. Compliance diminishes the chance of data breaches, improves an organization's overall cybersecurity posture, and gives stakeholders faith in the security procedures.

## 4.7 Continuous Security Monitoring and Improvement

Continuous security monitoring (CSM) helps organizations manage risk. It automates the monitoring of cyber threats, vulnerabilities, and information security policies (He et al, 2022). Traditional security solutions like firewalls and antivirus software are insufficient in today's digital environment to stop cyber attackers from using increasingly sophisticated approaches (Aslan et al., 2023). Throughout the world, governments are passing strict data protection legislation, enforcing the reporting of data breaches, and levying heavy fines for noncompliance. Companies now have larger attack surfaces due to outsourcing and subcontracting non-essential business tasks; therefore, careful monitoring is vital to reduce the risks of third and fourth parties. In order to enable proactive risk management and threat mitigation, CSM works by giving real-time insights into an organization's security posture. Organizations may effectively manage their cybersecurity risks by evaluating security measures, reviewing security-related data, and maintaining situational awareness (Renaud & Ophoff, 2021). The organization's total security resilience is improved by this data-driven strategy, which also helps with regulatory compliance. Proactive threat detection, constant risk assessment, and asset discovery are all part of implementing best practices for continuous security monitoring (Landoll, 2021).

## 4.8  Secure Supply Chain Management

Application systems are sometimes built from a patchwork of open-source libraries and components, and the idea of the software supply chain has undergone tremendous change in recent years. This method creates vulnerabilities outside an organization's immediate control, even if it allows quick development and implementation. Organizations inherit, by default, the complicatedness of the software supply chain (Kestilä, 2022). The software supply chain is vulnerable because it is interspersed, providing multiple ports of entry for attackers to exploit. These weaknesses affect people, technology, and procedures at several touchpoints. The commonness of security breaches originating from misconfigurations in cloud systems indicates the severe threats posed by infrastructure vulnerabilities. Similarly, malicious malware can be inserted or security measures compromised by taking advantage of software vulnerabilities, especially in open-source libraries and third-party tools. The risk environment is further worsened by vulnerabilities introduced by developers or by inefficient procedures like identity and access management (IAM) protocols (Olabanji et al., 2024). The reliance on external components exposes businesses to possible exploitation by hostile actors using common vulnerabilities and exposures (CVEs), even with the advantages of this approach, such as cost reduction and speedier development (Tuptuk & Hailes, 2018). Vulnerabilities can be proactively addressed by automating supply chain security through continuous integration and delivery (CI/CD) pipelines (Koopman, 2019). Many risk-averse organizations, including governments and organizations, need a software bill of materials (SBOM) that outlines all parts to solve this (Arora et al., 2022).

## 5. KEY PERFORMANCE INDICATORS (KPIS) FOR SOFTWARE SUPPLY CHAIN SECURITY

### 5.1. Vulnerability Resolution Rate (VRR)

This indicates how quickly and effectively vulnerabilities are fixed when they are discovered (Shah et al., 2019). The time that passes between the time an issue is discovered or reported and the time it is closed or fixed is known as the incident reaction time including the periods for acknowledgment, assignments, resolution, and closure (Serrano et al., 2023). In general, they are a more robust and effective security posture is indicated by a quicker incident reaction time. Prompt identification and reaction lessens the possible harm resulting from security breaches, shorten the time systems are unavailable, and improve an organization's capacity to adjust and bounce back from new risks (Serrano et al., 2023).

However, critics have noted that this approach does not proactively address vulnerabilities but focuses mainly on addressing them after they have been identified. This reactive approach may overlook systemic flaws within the supply chain can could cause recurring vulnerabilities. Besides, VRR may incentivize quick fixes at the expense of comprehensive remediation, causing ineffective or incomplete solutions. Besides, it may overlook the root causes of vulnerabilities in the supply chain (Hao et al., 2023).

### 5.2. Vulnerability Reopen Rate

Recurring vulnerabilities fixed in the same or a different asset point to a system configuration problem that requires careful examination (Jacobs et al., 2020). The vulnerability reopens rate measure provides information on how frequently vulnerabilities resurface due to shortcomings in patch management and vulnerability remediation procedures (Mansourov and Campara, 2011). This KPI shows how successful patch management and vulnerability repair procedures are. Reopening a fixed vulnerability often suggests serious flaws in the remediation procedure (Parker, 2023). Consequently, maintaining a robust security posture, ensuring that vulnerabilities are completely fixed, and stopping the recurrence of known security flaws in software systems depend on tracking and minimizing the Vulnerability Reopen Rate (Orebaugh and Pinkard, 2018). However, this metric has key limitations. First, it does not differentiate between critical and minor vulnerabilities, potentially causing a misinterpretation of the whole resilience level. It also incentivizes rapid vulnerability closures without identifying root causes, requiring businesses to conduct proactive and qualitative assessments to strengthen supply chain resilience (dos Santos and Nunes, 2018).

## 5.3. Frequent Security Evaluations and Audits

Resilience measurement requires regular security audits and assessments since they offer detailed information about how well security procedures are working. This includes:

## 5.4. Penetration Testing

In penetration testing, also known as pen testing, a cyber-security specialist looks for and attempts to take advantage of weaknesses in a computer system (Shah and Mehtre, 2016). To identify and illustrate the effects of system flaws on corporate operations, penetration testers employ the same instruments, strategies, and procedures as attackers. This assault simulation aims to find any vulnerabilities in a system's defenses that an attacker may exploit. This is because, as Wang et al. (2021) demonstrate, to find configuration errors, possible entry points for unauthorized access, and security control gaps, testers employ both automated tools and human methods.

However, according to Börstler et al. (2023), penetration testing has certain limitations. First, it only provides a point-in-time snapshot of security posture and may not fully capture post-assessment vulnerabilities. Continuous threat exploitation implies that some vulnerabilities may surface between tests, exposing systems. Besides, this requires skilled experts to effectively simulate attacks, making it costly and resource-intensive for some companies.

## CONCLUSION AND RECOMMENDATIONS

Software supply chain security combines risk management and standards for cybersecurity to help safeguard the software supply chain against possible vulnerabilities. Due to challenges ranging from the complexity of the supply chain, encompassing the complexity of supply chains, limited visibility into software origins, and diverse cybersecurity standards. Effective strategies include secure development practices, third-party dependency management, software compliance, zero-trust implementation, and stringent identity access management.

Cybersecurity laws and regulations are paramount in safeguarding organizations from the pervasive threats posed by cybercriminals. These advisory and legal frameworks are critical in establishing a robust defense against the sophisticated and ever-evolving tactics employed by malicious actors(Ejiofor et al.2023). Adherence to these guidelines and updates is crucial for enhancing the security posture of organizations. Regulatory bodies play a pivotal role in reminding private organizations of their responsibilities to secure customer data and maintain its privacy—even when there might be temptations to compromise security for increased profit or treat security as an afterthought. In the United States, for example, the Cybersecurity and Infrastructure Security Agency (CISA) issued advisory guidelines related to "Secure by Design" principles and the European Union issued regulatory guidelines - Cyber Resilience Act (CRA). Private organizations must ensure their development teams are well-versed in these guidelines and implement them across the board to foster a safer operational environment. By integrating these advisory/regulatory standards, organizations not only comply with legal requirements but also significantly bolster their defenses against cyber threats, thereby protecting their reputation and ensuring the trust of their customers.

International Journal on Soft Computing (IJSC) Vol.15, No.1/2, May 2024
Third-Party Libraries and Plug-Ins across Platforms. *Systems*, [online] 11(5), p.262. doi: https://doi.org/10.3390/systems11050262.

[42] Ivimey-Cook, E.R., Pick, J.L., Bairos-Novak, K.R., Antica Culina, Gould, E., Grainger, M., Marshall, B.M., Moreau, D., Paquet, M., Raphaël Royauté, Sánchez-Tójar, A., Silva, I. and Windecker, S.M. (2023). Implementing code review in the scientific workflow: Insights from ecology and evolutionary biology. *Journal of Evolutionary Biology*, 36(10), pp.1347–1356. doi: https://doi.org/10.1111/jeb.14230.

[43] Jacobs, J., Romanosky, S., Adjerid, I. and Baker, W. (2020). Improving vulnerability remediation through better exploit prediction. *Journal of Cybersecurity*, [online] 6(1). doi: https://doi.org/10.1093/cybsec/tyaa015.

[44] Jbair, M., Ahmad, B., Maple, C., & Harrison, R. (2022). Threat modelling for industrial cyber physical systems in the era of smart manufacturing. *Computers in Industry*, 137, 103611.

[45] Keskin, O.F., Caramancion, K.M., Tatar, I., Raza, O. and Tatar, U. (2021). Cyber Third-Party Risk Management: A Comparison of Non-Intrusive Risk Scoring Reports. *Electronics*, [online] 10(10), p.1168. doi: https://doi.org/10.3390/electronics10101168.

[46] Kestilä, R. (2022). *Acknowledging the risks of open source dependencies to software supply chain security* (Master's thesis).

[47] Koopman, M. (2019). *A framework for detecting and preventing security vulnerabilities in continuous integration/continuous delivery pipelines* (Master's thesis, University of Twente).

[48] Kumar, S. and Nottestad, D.A. (2012). Supply chain analysis methodology – Leveraging optimization and simulation software. *OR Insight*, 26(2), pp.87–119. doi: https://doi.org/10.1057/ori.2012.10.

[49] Landoll, D. (2021). *The security risk assessment handbook: A complete guide for performing security risk assessments*. CRC press.

[50] Li, J., Maiti, A., & Fei, J. (2023). Features and Scope of Regulatory Technologies: Challenges and Opportunities with Industrial Internet of Things. *Future Internet*, 15(8), 256.

[51] Librantz, A.F.H., Costa, I., Spinola, M. de M., de Oliveira Neto, G.C. and Zerbinatti, L. (2020). Risk assessment in software supply chains using the Bayesian method. *International Journal of Production Research*, 59(22), pp.6758–6775. doi: https://doi.org/10.1080/00207543.2020.1825860.

[52] Linton, J.D., Boyson, S. and Aje, J. (2014). The challenge of cyber supply chain security to research and practice – An introduction. *Technovation*, [online] 34(7), pp.339–341. doi: https://doi.org/10.1016/j.technovation.2014.05.001.

[53] Mansourov, N. and Campara, D. (2011). *Chapter 7 - Vulnerability patterns as a new assurance content*. [online] ScienceDirect. Available at: https://www.sciencedirect.com/science/article/abs/pii/B9780123814142000075 [Accessed 4 Feb. 2024].

[54] Melara, M. S., Liu, D. H., & Freedman, M. J. (2019). Pyronia: Intra-Process Access Control for IoT Applications. *arXiv preprint arXiv:1903.01950*.

[55] Meng, N., Nagy, S., Yao, D., Zhuang, W., &Argoty, G. A. (2018, May). Secure coding practices in java: Challenges and vulnerabilities. In *Proceedings of the 40th International Conference on Software Engineering* (pp. 372-383).

[56] Mubarkoot, M., Altmann, J., Rasti-Barzoki, M., Egger, B. and Lee, H. (2022). Software Compliance Requirements, Factors, and Policies: a Systematic Literature Review. *Computers & Security*, p.102985. doi: https://doi.org/10.1016/j.cose.2022.102985.

[57] Nazir, S., Price, B., Surendra, N.C. and Kopp, K. (2022). Adapting agile development practices for hyper-agile environments: lessons learned from a COVID-19 emergency response research project. *Information Technology and Management*. doi: https://doi.org/10.1007/s10799-022-00370-y.

[58] Nissen and Sengupta (2006). Incorporating Software Agents into Supply Chains: Experimental Investigation with a Procurement Task. *MIS Quarterly*, 30(1), p.145. doi:https://doi.org/10.2307/25148721.

[59] Nyre-Yu, M., Gutzwiller, R.S. and Caldwell, B.S. (2019). Observing Cyber Security Incident Response: Qualitative Themes From Field Research. *Proceedings of the Human Factors and Ergonomics Society Annual Meeting*, 63(1), pp.437–441. doi: https://doi.org/10.1177/1071181319631016.

[60] Ogheneovo, E.E. (2014). On the Relationship between Software Complexity and Maintenance Costs. *Journal of Computer and Communications*, 02(14), pp.1–16. doi: https://doi.org/10.4236/jcc.2014.214001.

International Journal on Soft Computing (IJSC) Vol.15, No.1/2, May 2024[102] Xiong, W., &Lagerström, R. (2019). Threat modeling–A systematic literature review. *Computers & security*, *84*, 53-69.

[103] Yauney, J., Bartholomew, S.R. and Rich, P. (2021). A systematic review of 'Hour of Code' research. *Computer Science Education*, pp.1–33. doi: https://doi.org/10.1080/08993408.2021.2022362.

[104] Yeboah-Ofori, A. and Islam, S. (2019). Cyber Security Threat Modeling for Supply Chain Organizational Environments. *Future Internet*, 11(3), p.63. doi: https://doi.org/10.3390/fi11030063.

[105] Zhang, E. (2022). *What is Role-Based Access Control (RBAC)? Examples, Benefits, and More*. [online] Digital Guardian. Available at: https://www.digitalguardian.com/blog/what-role-based-access-control-rbac-examples-benefits-and-more. (Accessed 5th February 2024)

[106] Zhu, J., Xie, J., Lipford, H.R. and Chu, B. (2014). Supporting secure programming in web applications through interactive static analysis. *Journal of Advanced Research*, 5(4), pp.449–462. doi: https://doi.org/10.1016/j.jare.2013.11.006.

[107] https://www.techtarget.com/searchsecurity/news/252502216/Mandiant-Compromised-Colonial-Pipeline-password-was-reused

[108] https://content.reversinglabs.com/sscs-report/guidance-and-initiatives

[109] https://www.cisa.gov/sites/default/files/publications/ESF_SECURING_THE_SOFTWARE_SUPPLY_CHAIN_DEVELOPERS.PDF
18